\documentclass[final,twocolumn]{IEEEtran}
\usepackage{graphicx,subfigure,psfrag,epsfig,epsf,latexsym,hhline,amsmath,amssymb,amsthm,multirow,booktabs,rotating}

\begin{document}

\title{Joint Compute and Forward for the Two Way Relay Channel with Spatially Coupled LDPC Codes\thanks{This work was supported by the National Science Foundation under Grant CCF 0830696.}}
\author{Brett Hern and Krishna Narayanan\\
Department of Electrical and Computer Engineering \\
Texas A\&M University\\
College Station\\
TX 77843, U.S.A}
\maketitle

\begin{abstract}

We consider the design and analysis of coding schemes for the binary input two way relay channel with erasure noise. We are particularly interested in reliable physical layer network coding in which the relay performs perfect error correction prior to forwarding messages. The best known achievable rates for this problem can be achieved through either decode and forward or compute and forward relaying. We consider a decoding paradigm called joint compute and forward which we numerically show can achieve the best of these rates with a single encoder and decoder. This is accomplished by deriving the exact performance of a message passing decoder based on joint compute and forward for spatially coupled LDPC ensembles.


\end{abstract}


\section{Introduction}

\begin{figure}
\centering
\includegraphics[width=3.5in]{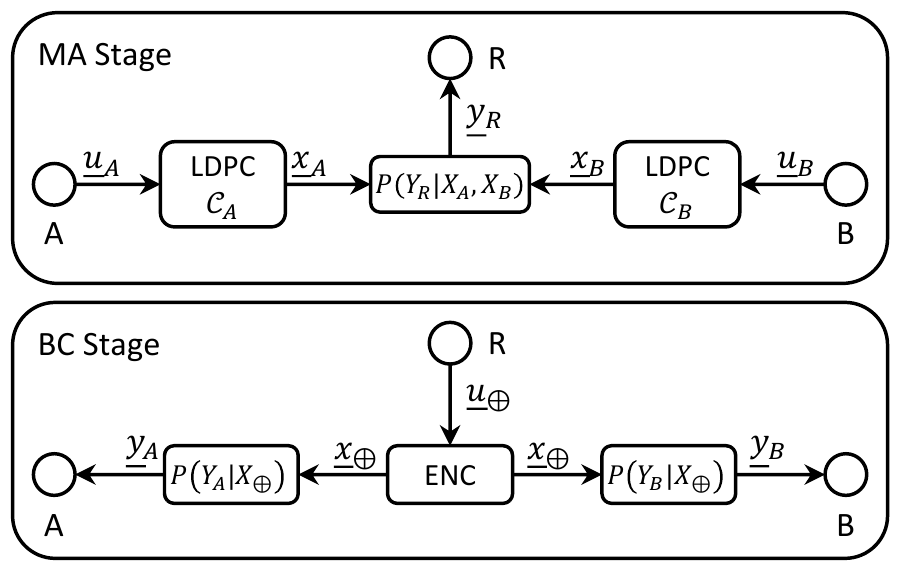}
\caption{System model showing a of two-way relay channel with PLNC.}
\label{fig:TWEChannel}
\end{figure}

We consider the two way relaying channel in which node $A$ has data to send to node $B$ and vice versa. The relay $R$ is included to assist in this communication, and there is no direct link between nodes $A$ and $B$. Let $\underline{u}_A,\underline{u}_B\in\{0,1\}^K$ be the binary input message sequences\footnote{We use underlined vectors like $\underline{x}$ to refer to symbol sequences, and we use $x$ or $x[n]$ to refer to an arbitrary element or the $n$th element of $\underline{x}$, respectively.} which are encoded into length $N$ codewords $\underline{x}_A\in\mathcal{C}_A$ and $\underline{x}_B\in\mathcal{C}_B$. The relay observes the output of a memoryless multiple access channel (MAC) $P(Y_R|X_A,X_B)$. The objective of the relay is to reliably decode the message $\underline{u}_{\oplus}=\underline{u}_A\oplus\underline{u}_B$ which is encoded and broadcast to nodes A and B. This is shown in Fig. \ref{fig:TWEChannel}. We define the achievable computation rate $\mathcal{R}$ as the largest $\frac{K}{N}$ such that $\underline{u}_{\oplus}$ can be decoded reliably in the usual information-theoretic sense. We focus specifically on the achievable computation rates of Low Density Parity Check (LDPC) code ensembles and message passing decoding.

A natural solution to the aforementioned problem is to treat the channel as a multiple-access (MAC) channel. Then, two independent codes $\mathcal{C}_A$, $\mathcal{C}_B$ can be used such that $\underline{x}_A \in \mathcal{C}_A$ and $\underline{x}_B \in \mathcal{C}_B$ can be fully decoded at the relay. The relay can subsequently recover $\underline{u}_{\oplus}$. This scheme, called decode and forward (DF), can achieve all rates subject to \cite{cover2006elements}
\begin{align} \label{eq:R_DF_Binary}
\mathcal{R}_{DF} < \min\{&\frac{1}{2}I(Y_R;X_A,X_B), I(Y_R;X_A|X_B), \nonumber \\
&I(Y_R;X_B|X_A)\}.
\end{align}
When the two codes $\mathcal{C}_A$ and $\mathcal{C}_B$ are identical linear codes, in \cite{hern2011multilevel}, we showed that the following rates are achievable
\begin{align} \label{eq:R_DF_Binary_SameCode}
\mathcal{R}'_{DF} < \min\{& \mathcal{R}_{DF},I(Y_R;X_A,X_B|X_{A\oplus B})\}.
\end{align}

A different strategy called compute and forward (CF) became popular in research because of the idea that it is wasteful for a relay to reliably decode each received message \cite{nazer2011reliable}, \cite{DBLP:journals/corr/abs-0805-0012}. In CF, nodes A and B use the same linear code $\mathcal{C}$, and the relay directly computes
\begin{align} \label{eq:MAP_CF_Dec_Def}
\hat{\underline{x}}_{\oplus,CF}(\underline{y}_R) &= \underset{\underline{x}_{\oplus}\in\mathcal{C}}{\arg\max}\prod_{n=1}^{N}{P(y_R[n]|x_{\oplus}[n])}.
\nonumber \\
&=
\underset{\underline{x}_{\oplus}\in\mathcal{C}}{\arg\max} \sum_{\{\underline{x}_A,\underline{x}_B\in\{0,1\}^N|\underline{x}_{\oplus}=\underline{x}_A\oplus \underline{x}_B\}}{P(\underline{y}_R|\underline{x}_A,\underline{x}_B)}.
\end{align}
CF can achieve all rates subject to
\begin{equation} \label{eq:R_CF_Binary}
\mathcal{R}_{CF} < I(Y_R;X_{\oplus}).
\end{equation}
This decoding structure performs well when $P(Y_R|X_A,X_B)$ is matched to $X_{\oplus}$ in some sense \cite{nazer2007computation}. According to \cite{nazer2007computation}, $P(Y_R|X_A,X_B)$ is ideally matched to $X_{\oplus}$ if the channel satisfies $P(Y_R|X_A,X_B)=P(Y_R|X_A\oplus1,X_B\oplus1)=P(Y_R|X_{\oplus})$. 
A CF decoder assumes ideal matching which means that it effectively performs error correction on elementwise estimates $P(Y_R|X_{\oplus})$.


When $P(Y_R|X_A,X_B)$ and $X_{\oplus}$ are poorly matched, the information lost by operating on the estimates $P(Y_R|X_{\oplus})$ is significant (i.e. $P(Y_R|X_{\oplus})$ is not a sufficient statistic). Therefore several schemes which perform message passing decoding with larger message alphabets have been proposed \cite{zhang2009channel}, \cite{wubben2010generalized}, \cite{lu2011asynchronous}, \cite{lu2011optimal}. We refer to the decoding paradigm used in these papers as Joint Compute and Forward (JCF) and we note that this decoder attempts to solve the following problem
\begin{equation} \label{eq:MAP_JCF_Dec_Def}
\hat{\underline{x}}_{\oplus,JCF}(\underline{y}_R) = \underset{\underline{x}_{\oplus}\in\mathcal{C}}{\arg\max} \sum_{\{\underline{x}_A,\underline{x}_B\in\mathcal{C} | \underline{x}_{\oplus}=\underline{x}_A\oplus \underline{x}_B\}}{P(\underline{y}_R|\underline{x}_A,\underline{x}_B)}.
\end{equation}
Note that the summation is taken over the set of codewords $\mathcal{C}$ instead of the set $\{0,1\}^N$ as in \eqref{eq:MAP_CF_Dec_Def}. Comparing (\ref{eq:MAP_CF_Dec_Def}) and (\ref{eq:MAP_JCF_Dec_Def}), it is clear that the JCF decoder is better than the CF decoder. Indeed, there are simulation results in \cite{zhang2009channel}, \cite{wubben2010generalized}, \cite{lu2011asynchronous}, \cite{lu2011optimal} which indicate that with JCF, higher computation rates than those achievable with the CF strategy in \eqref{eq:R_CF_Binary} can be achieved for certain channel parameters. However, the results in \cite{zhang2009channel}, \cite{wubben2010generalized}, \cite{lu2011asynchronous}, \cite{lu2011optimal} are based only on simulations and provide little insight into the operation of the JCF decoder. Particularly, the relationship between $\mathcal{R}_{DF},~\mathcal{R}_{CF}$, and the rates achievable with JCF is not clear. This is because it is difficult to develop analytical results for the JCF decoder.

In this paper, we introduce a class of two-way erasure multiple access channels (TWEMAC) for which we can exactly analyze the performance of the JCF message passing decoder. Specifically, we derive the exact density evolution equations and compute the corresponding thresholds for LDPC code ensembles with JCF message passing decoding. For a given set of channel parameters, we numerically show that spatially-coupled LDPC codes when decoded in this way can achieve all rates subject to
\begin{equation} \label{eq:R_JCF}
\mathcal{R}_{JCF} < \max\{\mathcal{R}_{DF},\mathcal{R}_{CF}\}.
\end{equation}
For any binary input memoryless MAC channel, it can be shown that $\max\{\mathcal{R}'_{DF},\mathcal{R}_{CF}\}=\max\{\mathcal{R}_{DF},\mathcal{R}_{CF}\}$. The proof of this must be omitted for space. This means that the JCF decoder naturally performs as well as the better of the DF and CF schemes.

Finally, we show that the class of punctured spatially-coupled LDPC codes can be used to obtain a rate of $\max\{\mathcal{R}_{DF},\mathcal{R}_{CF}\}$ over the entire range of channel parameters, i.e., it is not required to use different degree distributions of LDPC codes for the different rates and channel parameters. It suffices to use a single spatially coupled LDPC code and puncture this code until the rate is close to $\max\{\mathcal{R}_{DF},\mathcal{R}_{CF}\}$ for given channel parameters.

\section{Channel Model}
We introduce the TWEMAC channel model and characterize the performance of iterative decoding of LDPC ensembles over the TWEMAC. We introduced a special case of this channel very recently in \cite{Khis1207:Modulo}. A TWEMAC is randomly in one of five states $\tau\in\mathcal{T}=\{1,...,5\}$ during each channel use. The value of $y_R$ is a deterministic function of $x_A,x_B$ and the state $\tau$. In this paper, the probability $p_{\tau}$ that the channel is in state $\tau$ is parameterized by an erasure probability $\epsilon\in[0,1]$. Thus the channel is defined
\begin{equation} \label{eq:TWEMAC_Def}
Y_R=\left\{\begin{array}{lll}
(E,E) & \textrm{with probability} & p_1(\epsilon) \\
(X_A,E) & \textrm{with probability} & p_2(\epsilon) \\
(E,X_B) & \textrm{with probability} & p_3(\epsilon) \\
(X_A\oplus X_B) & \textrm{with probability} & p_4(\epsilon) \\
(X_A,X_B) & \textrm{with probability} & p_5(\epsilon) \\
\end{array}\right.
\end{equation}
where $E$ denotes an erasure and $\sum_{i=1}^5 p_i(\epsilon) = 1 ~\forall~\epsilon\in [0,1]$. The state of the channel is assumed to be known to the relay but unknown at nodes A and B.

Since we are interested in message passing decoding, it is useful to refer to the messages from the channel as having a message type from the set $\mathcal{T}$. Then messages passed along the edges in the decoder will also have a type from the set $\mathcal{T}$. We define the type distribution vector for a given channel according to
\begin{equation} \label{eq:PchDef}
\underline{P}_{ch}(\epsilon) = \left[p_1(\epsilon), p_2(\epsilon), p_3(\epsilon), p_4(\epsilon), p_5(\epsilon)\right]^T
\end{equation}
where the subscript $ch$ associates this type distribution with the channel output. The class of TWEMACs has the advantage that the probabilities $p_\tau,\tau\in\mathcal{T}$ can be chosen in order to mimic or isolate many characteristics of wireless channels.

For this paper, we will present numerical results for the TWEMAC parameterized by
\begin{equation} \label{eq:PchPrimary}
\underline{P}_{ch}(\epsilon) = \left[\epsilon^2, (1-\epsilon)\epsilon, \epsilon(1-\epsilon), (1-\epsilon)^2, 0 \right]^T.
\end{equation}
This channel is generated if we assume that $x_A$ and $x_B$ are erased with probability $\epsilon$ independently. When neither symbol is erased, the relay observes the type $4$ message $x_{\oplus}$. When $x_A$ ($x_B$) is erased and $x_B$ ($x_A$) is not, we receive a message of type $3$ (type $2$). This channel mimics the effects of deep fade events for a wireless channel. The channel coefficients can be either $1$ or $0$, and when both channels are strong, the matching for the channel is ideal. We fix $p_5(\epsilon)=0$ so that the change in matching quality with $\epsilon$ is more pronounced.

\section{CF and JCF Message Passing Decoders}
We consider the case where $\underline{x}_A,\underline{x}_B\in\mathcal{C}$ so that the codewords are members of {\em identical linear LDPC} codes. By the linearity of $\mathcal{C}$, we have $\underline{x}_{\oplus}\in\mathcal{C}$. The constraints associated with $\underline{x}_{\oplus}\in\mathcal{C}$ can be described by a Tanner graph with variable nodes $x_{\oplus}[n]\in\{0,1\}$.  A decoder which directly decodes $\underline{x}_{\oplus}$ by passing messages in this Tanner graph is a CF message passing decoder.

\begin{figure}
\centering
\includegraphics[width=3.5in]{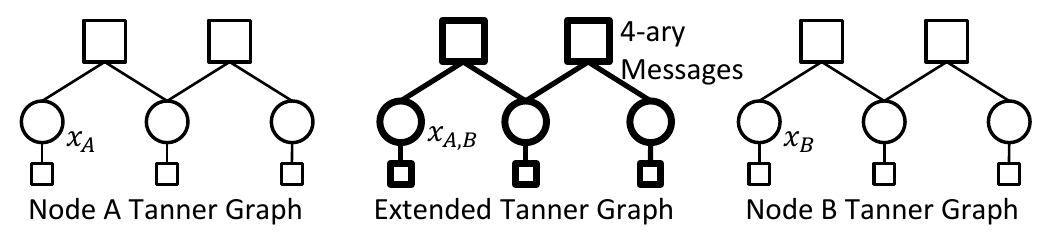}
\caption{ETG when nodes A and B use a $(3,1)$ repetition code.}
\label{fig:ExtendedTannerGraph}
\end{figure}

In addition to the $\underline{x}_{\oplus}\in\mathcal{C}$ constraints, a JCF decoder uses the constraints associated with $\underline{x}_A\in\mathcal{C}$ and $\underline{x}_B\in\mathcal{C}$ for decoding. All three of these constraint sets can be jointly expressed by an Extended Tanner Graph (ETG) as shown in Fig. \ref{fig:ExtendedTannerGraph}. Therefore, the decoder which decodes $\underline{x}_{\oplus}$ by passing messages on this ETG is a JCF message passing decoder. In the ETG, each variable node $x_{A,B}[n]=[x_A[n],x_B[n]]^T,~n\in\{1,...,N\}$ takes a value from $\{0,1\}^2$. The edges carry $4-ary$ messages which are local estimates of $x_{A,B}[n]$ of the form
\begin{equation} \label{eq:ETG_Messages}
\mu_n = \left[\begin{array}{c}
P(X_{A,B}[n] = [00]^T|\underline{Y}_R) \\
P(X_{A,B}[n] = [01]^T|\underline{Y}_R) \\
P(X_{A,B}[n] = [10]^T|\underline{Y}_R) \\
P(X_{A,B}[n] = [11]^T|\underline{Y}_R)
\end{array}\right].
\end{equation}
A rigorous derivation of the message processing rules for the JCF message passing decoder must be omitted for space, however these rules follow directly from \cite{richardson2008modern}.

While the messages passed along the edges in the ETG are 4-ary messages, to characterize the performance of an LDPC ensemble with a JCF message passing decoder, it is sufficient to think of these messages as being one of the 5 types from the set $\mathcal{T}$. For example, if $\mu_n = [\frac12 \ \frac12 \ 0 \ 0]^T$, this essentially corresponds to $X_A[n]$ being perfectly known and $X_B[n]$ being unknown or, equivalently, erased. Hence, we can think of this as being a message of type 2. In this paper, we use the latter representation as this will make it more convenient to analyze the decoder.

The input output relationship for message types $\mathcal{T}$ at the variable and check nodes can be obtained from the message processing rules for the $4-ary$ estimates of each $x_{A,B}[n]$ as follows.

\begin{table} [h]
\centering
\begin{tabular}{| c | c || c | c | c | c | c |}
\hline
\multicolumn{2}{|c|}{\multirow{2}{*}{VAR}} & \multicolumn{5}{|c|}{$\tau_1^{in}$} \\ \cline{3-7}
\multicolumn{2}{|c|}{} &  1   &   2   &   3   &   4   &   5   \\ \hline \hline
\multicolumn{1}{|c|}{\multirow{5}{*}{$\tau_2^{in}$}}
&   1                            &   1   &   2   &   3   &   4   &   5   \\ \cline{2-7}
&   2                            &   2   &   2   &   5   &   5   &   5   \\ \cline{2-7}
&   3                            &   3   &   5   &   3   &   5   &   5   \\ \cline{2-7}
&   4                            &   4   &   5   &   5   &   4   &   5   \\ \cline{2-7}
&   5                            &   5   &   5   &   5   &   5   &   5   \\ \hline
\end{tabular}
\caption{Variable node operator table with respect to types $\mathcal{T}$.}
\label{tab:VarOpTypes}
\end{table}

Consider a degree $d_v$ variable node which is connected to $d_v$ check nodes and one function node associated with the channel observation. Thus, each output message from a degree $d_v$ variable node is a function of $d_v$ input messages. For a variable node of degree $d_v=2$, the type of the outgoing message $\tau^{out}$ is a function of the types of the incoming messages, namely $\tau_1$ and $\tau_2$. Let this function be denoted by $VAR$, i.e., $\tau^{out} = VAR(\tau_1,\tau_2)$. The function $VAR$ is defined in Table \ref{tab:VarOpTypes}. For a variable node of degree $d_v$, if the input messages have types $\tau_1,...,\tau_{d_v}$, then the output message type is found by recursively applying $VAR(\cdot,\cdot)$ according to
\begin{equation}
\tau^{out} = VAR(\tau_1,VAR(\tau_2,...VAR(\tau_{d_v-1},\tau_{d_v}))).
\end{equation}

A degree $d_c$ check node is connected to $d_c$ variable nodes. Thus, each output message for a degree $d_c$ check node is a function of $d_c-1$ input messages. For a check node of degree $d_c=3$, the type update operation $CHK(\tau_1^{in},\tau_2^{in})$ is defined in Table \ref{tab:ChkOpTypes}. If the input messages to a check node of degree $d_c$ have types $\tau_1,...,\tau_{d_c-1}$, then the output message type is found by recursively applying $CHK(\cdot,\cdot)$ as
\begin{equation}
\tau^{out} = CHK(\tau_1,CHK(\tau_2,...CHK(\tau_{d_c-2},\tau_{d_c-1}))).
\end{equation}

\begin{table} [h]
\centering
\begin{tabular}{| c | c || c | c | c | c | c |}
\hline
\multicolumn{2}{|c|}{\multirow{2}{*}{CHK}} & \multicolumn{5}{|c|}{$\tau_1^{in}$} \\ \cline{3-7}
\multicolumn{2}{|c|}{} &  1   &   2   &   3   &   4   &   5   \\ \hline \hline
\multicolumn{1}{|c|}{\multirow{5}{*}{$\tau_2^{in}$}}
&   1                            &   1   &   1   &   1   &   1   &   1   \\ \cline{2-7}
&   2                            &   1   &   2   &   1   &   1   &   2   \\ \cline{2-7}
&   3                            &   1   &   1   &   3   &   1   &   3   \\ \cline{2-7}
&   4                            &   1   &   1   &   1   &   4   &   4   \\ \cline{2-7}
&   5                            &   1   &   2   &   3   &   4   &   5   \\ \cline{1-7}
\end{tabular}
\caption{Check node operator table with respect to types $\mathcal{T}$.}
\label{tab:ChkOpTypes}
\end{table}

\section{Type Distribution Evolution}
For the normal binary erasure channel, the asymptotic (in length) performance of an LDPC code or ensemble is often analyzed using the idea of density evolution \cite{richardson2008modern}. The main idea in density evolution is to compute the probability that a message from a variable node to check node (or check node to variable node) is erased during the $l$th iteration. In this section, we characterize the performance of LDPC ensembles using JCF message passing decoding for TWEMACs. Analogous to tracking the probability of erasure we will track the distribution of message types sent on a randomly chosen edge. Hence, we refer to this performance characterization as type distribution evolution.

To be consistent with much of the literature, each variable node is initialized (iteration 0) with the observation from the channel and forwards this message to all connected check nodes. The check nodes process these messages and forward their estimates back to the variable nodes. This process continues until the iterations reach a fixed point. As in \cite{richardson2008modern}, we work under the assumption that, for a finite number of iterations and for some $N$ large enough, the local graph around any variable node is a tree with high probability. This allows us to assume that the type distribution for each input message to a variable or check node is independent from the other input messages.

We have defined the input output relationship for variable and check nodes on the set $\mathcal{T}$. Here, these rules are extended to the set of $5-ary$ type distributions. Similar to \eqref{eq:PchDef}, we define the type distribution of messages from the variable to check (check to variable) nodes during iteration $\ell$ as $\underline{P}_{vc}^{(\ell)}$ ($\underline{P}_{cv}^{(\ell)}$). We will index these pmfs according to
\begin{align}
\underline{P}_{vc}^{(\ell)} &= \left[P_{vc}^{1,(\ell)},P_{vc}^{2,(\ell)},P_{vc}^{3,(\ell)},P_{vc}^{4,(\ell)},P_{vc}^{5,(\ell)}\right]^T \nonumber \\
\underline{P}_{cv}^{(\ell)} &= \left[P_{cv}^{1,(\ell)},P_{cv}^{2,(\ell)},P_{cv}^{3,(\ell)},P_{cv}^{4,(\ell)},P_{cv}^{5,(\ell)}\right]^T.
\end{align}

\begin{figure}
\centering
\includegraphics[width=3in]{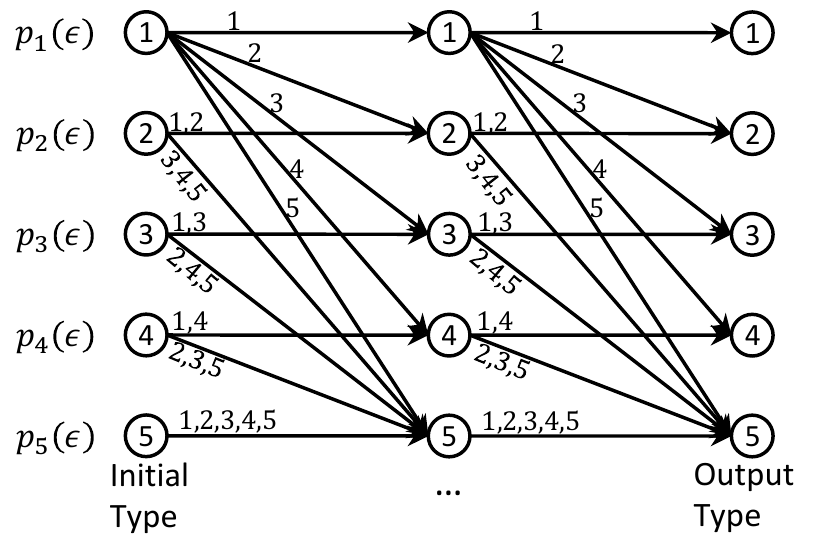}
\caption{Trellis diagram type distribution update for a degree 3 variable node.}
\label{fig:VarNodeTrellis}
\end{figure}

It is helpful to think of the variable node operation on the set of type distributions in terms of the probabilities of an initial state, given by $\underline{P}_{ch}$, and a subsequent transition probability, determined by $\underline{P}_{cv}^{(\ell)}$ and Table \ref{tab:VarOpTypes}. This is illustrated for a degree $d_v=3$ variable node in Fig. \ref{fig:VarNodeTrellis}. The distribution for the initial state is given by the channel type distribution $\underline{P}_{ch}$. Then, there are two input messages from check nodes whose type distributions are applied in the form of transitions through the trellis. These transitions are labeled according to the message types $\tau\in\mathcal{T}$ which correspond to the labeled transition. The probability associated with each transition is therefore the sum of the probabilities of the corresponding types (e.g. the probability of transition from type $3$ to $5$ is $P_{cv}^{2,(\ell)}+P_{cv}^{4,(\ell)}+P_{cv}^{5,(\ell)}$).

It is efficient to express this type distribution update operation in matrix form. For a degree $2$ variable node, we have
\begin{equation}
\underline{P}_{vc}^{(\ell)} = \mathbf{P}_{cv}^{(\ell)}\underline{P}_{ch}.
\end{equation}
The $(i,j)_{th}$ element of the matrix $\mathbf{P}_{cv}^{(\ell)}$ is defined
\begin{equation}
\left[\mathbf{P}_{cv}^{(\ell)}\right]_{i,j} = P(j\rightarrow i) = \sum_{\{k\in\mathcal{T} ~|~ i=VAR(k,j)\}} \underline{P}_{cv}^{k,(\ell)}
\end{equation}
where the elements of the summation follow the rules given in Table \ref{tab:VarOpTypes}. The notation $P(j\rightarrow i)$ refers to the probability of a transition from type $j$ to type $i$ as depicted in Fig. \ref{fig:VarNodeTrellis}.

This is equivalent to
\begin{equation}
\mathbf{P}_{cv}^{(\ell)} = \left[\begin{array}{ccccc}
P_{cv}^{1,(\ell)}   &   0   &   0   &   0   &   0                                 \\
P_{cv}^{2,(\ell)}   &   P_{cv}^{1+2,(\ell)}   &   0   &   0   &   0     \\
P_{cv}^{3,(\ell)}   &   0   &   P_{cv}^{1+3,(\ell)}   &   0   &   0     \\
P_{cv}^{4,(\ell)}   &   0   &   0   &   P_{cv}^{1+4,(\ell)}   &   0     \\
P_{cv}^{5,(\ell)}   &   P_{cv}^{3+4+5,(\ell)}   &   P_{cv}^{2+4+5,(\ell)}   &   P_{cv}^{2+3+5,(\ell)}   &   1     \\
\end{array}\right] \nonumber
\end{equation}
where the shorthand $P_{cv}^{i+j,(\ell)}=P_{cv}^{i,(\ell)}+P_{cv}^{j,(\ell)}$.

For a variable node of degree $d_v\geq 2$, the type distribution update equation can be stated as
\begin{align}
\underline{P}_{vc}^{(\ell)} = \left(\mathbf{P}_{cv}^{(\ell)}\right)^{d_v-1}\underline{P}_{ch}.
\end{align}
To see this, note that similar to multiplying $\mathbf{P}_{cv}^{(\ell)}\underline{P}_{ch}$, the matrix element $\left[\mathbf{P}_{cv}^{(\ell)}\mathbf{P}_{cv}^{(\ell)}\right]_{i,k}$ is defined by
\begin{align}
\left[\mathbf{P}_{cv}^{(\ell)}\mathbf{P}_{cv}^{(\ell)}\right]_{i,k} &= \sum_{j}P(k\rightarrow j)P(j\rightarrow i) \nonumber \\
&= \sum_{j}P(k\rightarrow j\rightarrow i)
\end{align}

\begin{figure}
\centering
\includegraphics[width=3in]{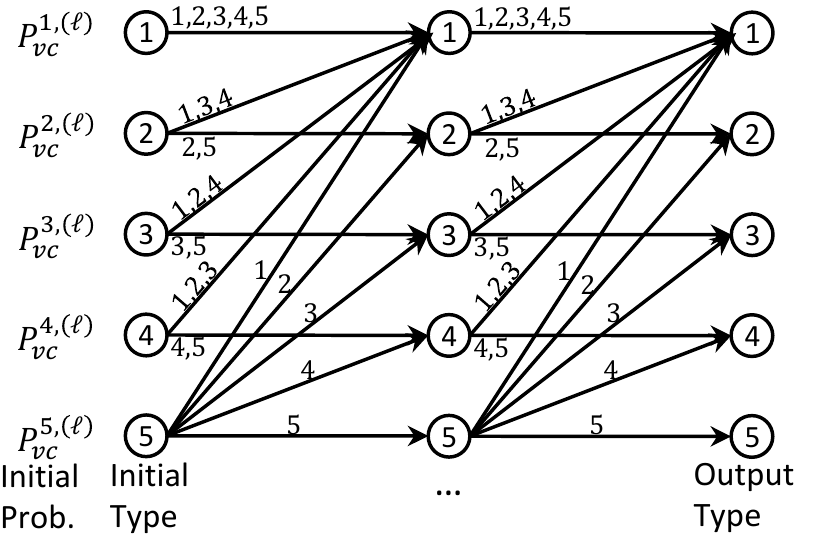}
\caption{Trellis diagram type distribution update for a degree 4 check node.}
\label{fig:ChkNodeTrellis}
\end{figure}

The check node type distribution update operation is similar. The trellis structure for a degree $4$ check node is depicted in Fig. \ref{fig:ChkNodeTrellis} where the transitions are arranged according to Table \ref{tab:ChkOpTypes}. The state transition matrix $\mathbf{P}_{vc}^{(\ell)}$ is defined
\begin{equation}
\mathbf{P}_{vc}^{(\ell)} = \left[\begin{array}{ccccc}
1   &   P_{vc}^{1+3+4,(\ell)}   &   P_{vc}^{1+2+4,(\ell)}   &   P_{vc}^{1+2+3,(\ell)}   &   P_{vc}^{1,(\ell)}                                 \\
0   &   P_{vc}^{2+5,(\ell)}   &   0   &   0   &   P_{vc}^{2,(\ell)}     \\
0   &   0   &   P_{vc}^{3+5,(\ell)}   &   0   &   P_{vc}^{3,(\ell)}     \\
0   &   0   &   0   &   P_{vc}^{4+5,(\ell)}   &   P_{vc}^{4,(\ell)}     \\
0   &   0   &   0   &   0   &   P_{vc}^{5,(\ell)}     \\
\end{array}\right]. \nonumber
\end{equation}
The update equation for a degree $d_c\geq2$ check node is therefore given by
\begin{equation}
\underline{P}_{cv}^{(\ell+1)} = \left(\mathbf{P}_{vc}^{(\ell)}\right)^{d_c-2}\underline{P}_{vc}^{(\ell)}.
\end{equation}
The state transition matrix $\mathbf{P}_{vc}^{(\ell)}$ is raised to the power $d_c-2$ because one of the input edges is forms the probabilities for the initial types. The remaining $d_c-2$ input edges are applied in the form of the state transition matrix (i.e. they form the transition probabilities in the trellis of Fig. \ref{fig:ChkNodeTrellis}).

\subsection{Type Distribution Evolution for Regular Ensembles}
For a $(d_v,d_c)$ regular ensemble, we assume that the incoming messages to a variable or check node are distributed according to $\underline{P}_{cv}^{(\ell)}$ or $\underline{P}_{vc}^{(\ell)}$ respectively. Since the first message from the variable to check nodes is the message from the channel, the type distribution evolution is initialized by
\begin{align}
\underline{P}_{vc}^{(0)} &= \underline{P}_{ch}\nonumber \\
\underline{P}_{cv}^{(1)} &= \left(\mathbf{P}_{vc}^{(0)}\right)^{d_c-2}\underline{P}_{vc}^{(0)}.
\end{align}
For every iteration $\ell\geq 2$, the type distribution evolution is fully characterized by
\begin{equation}
\underline{P}_{cv}^{(\ell)} = \left(\mathbf{P}_{vc}^{(\ell-1)}\right)^{d_c-2}
\underbrace{\left(\mathbf{P}_{cv}^{(\ell-1)}\right)^{d_v-1}\underline{P}_{ch}}_{\underline{P}_{vc}^{(\ell-1)}}.
\end{equation}

If we allow the decoder to complete $\ell_{max}$ iterations, the type distribution at the output of the decoder is
\begin{equation} \label{eq:POutDefReg}
\underline{P}_{out}^{(\ell_{max})} = \left(\mathbf{P}_{cv}^{(\ell_{max})}\right)^{d_v} \underline{P}_{ch}.
\end{equation}
The objective of the relay is to recover the codeword $\underline{x}_{\oplus}$. Therefore, the bitwise probability of successful computation after  $\ell_{max}$ iterations is defined by
\begin{equation} \label{eq:PDecDefReg}
P_{dec} = P_{out}^{4,(\ell_{max})} + P_{out}^{5,(\ell_{max})}.
\end{equation}
This is because the value of $x_{\oplus}$ is known if the decoder output is either type $4$ or $5$. We define $\epsilon_{thesh}$ as the largest $\epsilon$ such that reliable decoding is possible for an LDPC ensemble (i.e. $P_{dec}\stackrel{\ell_{max}\rightarrow\infty}{\xrightarrow{\hspace*{1cm}}}1$).

\subsection{Type Distribution Evolution Spatially Coupled Codes}
In this section, we characterize the performance of spatially coupled LDPC codes. Since node A and B each use the same spatially coupled LDPC code, the ETG used for decoding at the relay is also spatially coupled. The common method to achieve the capacity of a given channel is to find some distribution on the variable and check node degrees such that a message passing decoder can reliably decode at rates near the channel capacity. It has recently been discovered that for spatially coupled LDPC ensembles a message passing decoder can nearly achieve the performance of an optimal decoder \cite{kudekar2011threshold}, \cite{kudekar2011spatially}. This holds for a large class of binary input channels \cite{yedla2011universal}, \cite{kudekar2012spatially}.

Therefore, we extend our type distribution evolution analysis to the $(d_v,d_c,L,w)$ ensemble which is defined rigorously in \cite{kudekar2011spatially}. In this ensemble, $M$ variable nodes are placed in each position $i\in\{-L,...,L\}$ so that the codeword length is $N=(2L+1)M$. Each variable and check node has degree $d_v$ and $d_c$ respectively. There are $M d_v$ edges in each position which requires that $\frac{d_v}{d_c}M$ check nodes be placed at each position. These check nodes are placed at positions $i\in\{-L,...,L+w-1\}$. The ensemble is constructed by uniformly and independently connecting the $d_v$ ($d_c$) edges from a variable (check) node at position $i$ to check (variable) nodes at positions $\{i,...,i+w-1\}$ ($\{i-w+1,...,i\}$). All check node connections to variable nodes outside of positions $\{-L,...,L\}$ are connected to pseudo variable nodes whose value is fixed to $x_{A,B}=[00]^T$. This decreases the effective degree of these check nodes.

\begin{figure}
\centering
\includegraphics[width=3.5in]{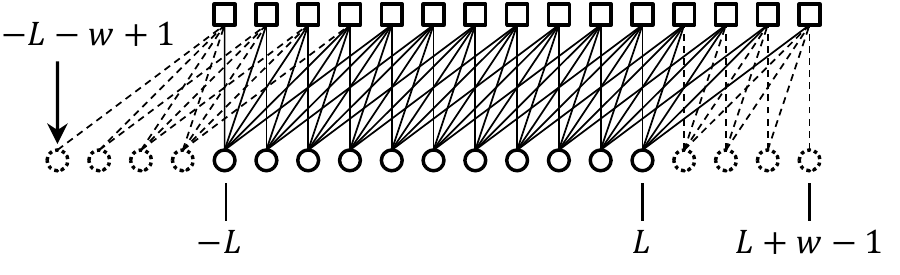}
\caption{Protograph diagram for a $(d_v,d_c,L=5,w=5)$ code ensemble.}
\label{fig:dvdcLwProtoDiagram}
\end{figure}

The nominal rate for a $(d_v,d_c,L,w)$ ensemble is defined
\begin{equation} \label{eq:R_dvdcLw}
\mathcal{R}(d_v,d_c,L,w) = \left(1-\frac{d_v}{d_c}\right) - \frac{d_v}{d_c}\frac{w+1-2\sum_{i=0}^{w}(\frac{i}{w})^{d_c}}{2L+1}.
\end{equation}
This is a lower bound on the actual rate, however, it approaches the rate of the regular ensemble $\mathcal{R}(d_v,d_c) = 1-\frac{d_v}{d_c}$ as $L\rightarrow\infty$ for a fixed $w$. This ensemble is primarily useful for information theoretic proofs. A protograph diagram for a $(d_v,d_c,L=5,w=5)$ ensemble is shown in Fig. \ref{fig:dvdcLwProtoDiagram} with variable and check nodes depicted by circles and squares respectively. The dashed edges depict possible connections to pseudo variable nodes which will always give a message of type $5$. Note that each edge in Fig. \ref{fig:dvdcLwProtoDiagram} represents a possible edge for a node at the given position. Particularly, a specific check node at position $-L$ might only have connections to pseudo variable nodes. 

To characterize the performance of this ensemble, we need to track the type distribution for the nodes in each position separately. We define the type distribution for messages from the output of a variable node at position $i$ to any check node as $\underline{P}_{vc,i}^{(\ell)}$. We define the {\em effective} input message to a check node at position $i$ as $\underline{\widetilde{P}}_{vc,i}^{(\ell)}$. We similarly define $\underline{P}_{cv,i}^{(\ell)}$ and $\underline{\widetilde{P}}_{cv,i}^{(\ell)}$. The effective input messages are calculated by
\begin{align} \label{eq:PEffDef}
\underline{\widetilde{P}}_{cv,i}^{(\ell)} &= \frac{1}{w}\sum_{j=i}^{i+w-1}\underline{P}_{cv,j}^{(\ell)}~\forall~ i\in\{-L,...,L\} \nonumber \\
\underline{\widetilde{P}}_{vc,i}^{(\ell)} &= \frac{1}{w}\sum_{j=i-w+1}^{i}\underline{P}_{vc,j}^{(\ell)}~\forall~ i\in\{-L,...,L+w-1\}.
\end{align}

The type distribution evolution for the $(d_v,d_c,L,w)$ ensemble is initialized by
\begin{align} \label{eq:TDE_Init_dvdcLw}
&\underline{P}_{vc,i}^{(0)} = \left\{\begin{array}{cc}
\underline{P}_{ch} &\forall~ i\in\{-L,...,L\}  \\
\left[00001\right]^T &\forall~ i\not\in\{-L,...,L\}
\end{array}\right. .
\end{align}

Then, the type distribution evolution for this ensemble is fully characterized by
\begin{align} \label{eq:TDE_dvdcLw}
&\textrm{FOR } \ell=1,...,\ell_{max} \nonumber \\
&~~~\underline{P}_{cv,i}^{(\ell)} = \left(\mathbf{\widetilde{P}}_{vc,i}^{(\ell-1)}\right)^{d_c-2}\underline{\widetilde{P}}_{vc,i}^{(\ell-1)}
~,~ i\in\{-L,...,L+w-1\}
\nonumber \\
&~~~\underline{P}_{vc,i}^{(\ell)} = (\mathbf{\widetilde{P}}_{cv,i}^{(\ell)})^{d_v-1}\underline{P}_{ch}
~~~~~~,~ i\in\{-L,...,L\}
\end{align}
where \eqref{eq:PEffDef} is used to determine $\underline{\widetilde{P}}_{cv,i}^{(\ell)}$ and $\underline{\widetilde{P}}_{vc,i}^{(\ell)}$ as needed.
Similar to \eqref{eq:POutDefReg}, if we allow $\ell_{max}$ iterations, the output type distribution for a variable node at position i for this decoder is
\begin{equation}
\underline{P}_{out,i}^{(\ell_{max})} = \left(\mathbf{\widetilde{P}}_{cv,i}^{(\ell_{max})}\right)^{d_v} \underline{P}_{ch}.
\end{equation}
Similar to \eqref{eq:PDecDefReg}, the bitwise probability of successful computation for a variable at position $i$ is
\begin{equation}
P_{dec,i} = P_{out,i}^{4,(\ell_{max})} + P_{out,i}^{5,(\ell_{max})}.
\end{equation}
We define $\epsilon_{thresh}$ as the largest $\epsilon$ such that reliable decoding is possible for this LDPC ensemble (i.e. $\min_{i\in\{-L,...,L\}}P_{dec,i}\stackrel{\ell_{max}\rightarrow\infty}{\xrightarrow{\hspace*{1cm}}}1$). Notice, that we are already considering the case that $M$ is very large. This type distribution analysis defines the average performance of the ensemble of protographs defined by the parameters $(d_v,d_c,L,w)$.

\subsection{Results and Discussion}

\begin{figure}
\centering
\includegraphics[width=3.5in]{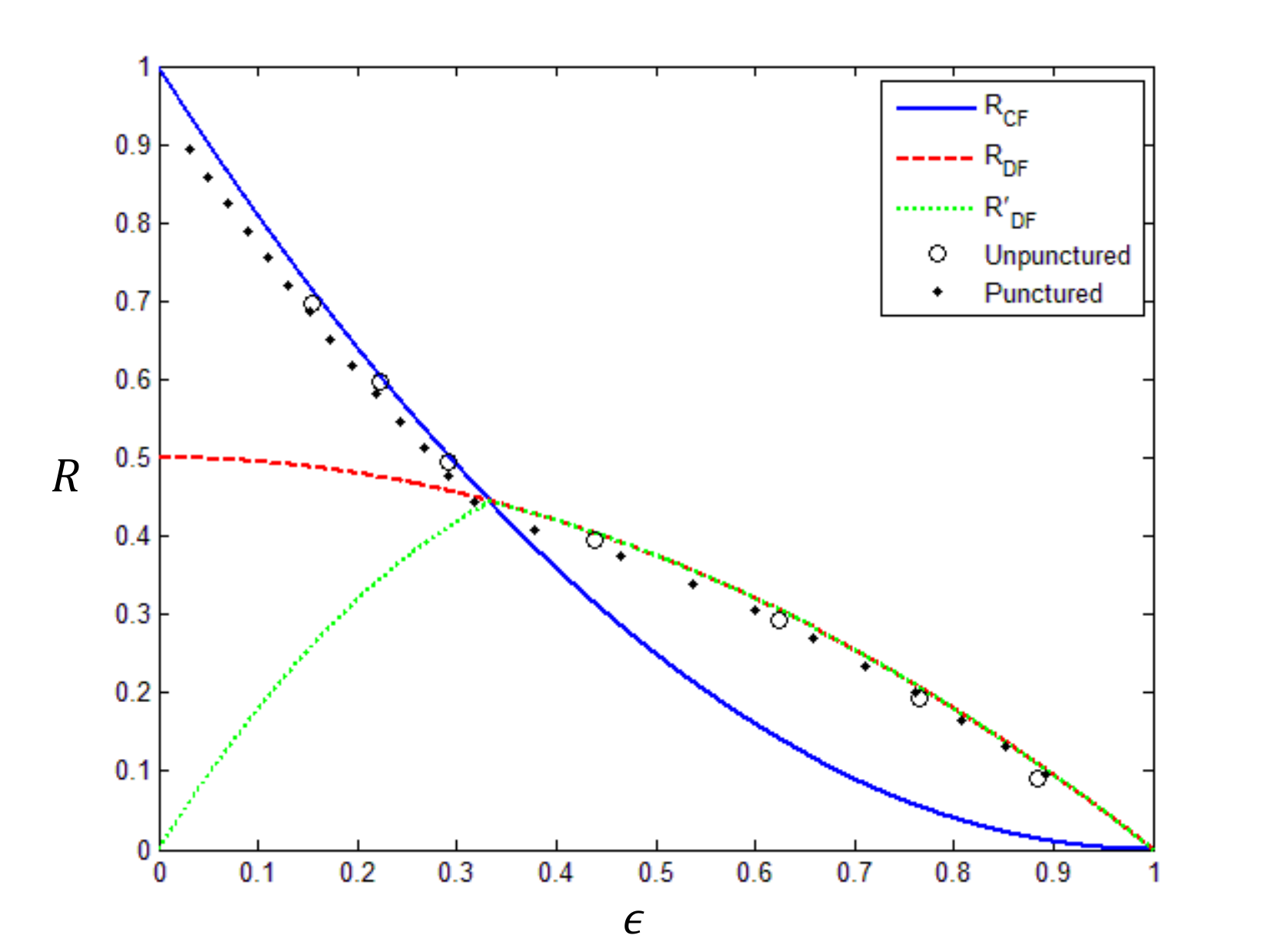}
\caption{$\mathcal{R}_{CF}$, $\mathcal{R}_{DF}$, $\mathcal{R}'_{DF}$, and numerical type distribution evolution results for $(d_v,d_c,L,w)=([3,...,9],10,4000,100)$ and $(d_v,d_c,L,w)=(9,10,10000,100)$ for un-punctured and punctured ensembles respectively.}
\label{fig:dvdcLwBoth}
\end{figure}

The rates $\mathcal{R}_{CF}$, $\mathcal{R}_{DF}$, and $\mathcal{R}'_{DF}$ are plotted along with numerical results for the thresholds obtained from type distribution evolution for a series of $(d_v,d_c,L,w)$ ensembles in Fig. \ref{fig:dvdcLwBoth}. Notice that $\mathcal{R}'_{DF}=\mathcal{R}_{DF}$ whenever $\mathcal{R}_{DF}\geq\mathcal{R}_{CF}$. The check degree is $d_c=10$. The protograph length parameter is $L=4,000$. The smoothing parameter is $w=100$. The variable node degree $d_v$ takes the values $d_v=3,...,9$ so that the performance can be characterized for several rates. The maximum number of iterations is $\ell_{max}=20,000$. Here, $\epsilon_{thresh}$ is the largest $\epsilon$ such that $min_{i\in\{-L,...,L\}}\{P_{dec,i}\}>1-10^{-5}$. The nominal rate \eqref{eq:R_dvdcLw} is plotted as a function of $\epsilon_{thresh}$. Notice that for each $\epsilon_{thresh}$, the achievable rate is closely upper bounded by the best of the $\mathcal{R}_{CF}$ and $\mathcal{R}_{DF}$ curves. This supports our claim that $\mathcal{R}_{JCF} = \max\{\mathcal{R}_{DF},\mathcal{R}_{CF}\}$. It is important to remember that the $(d_v,d_c,L,w)$ ensemble, and the very large $\ell_{max}$, is primarily useful for information theoretic analysis. More practical spatially coupled LDPC codes can be developed, and this will be the subject of future work.

It has been shown that spatially coupled ensembles can achieve near optimal decoding performance for some binary input MAC channels universally \cite{yedla2011universal}. This result coupled with \eqref{eq:R_JCF} suggests that it is practically possible to achieve all points on the $\max\{\mathcal{R}_{DF},\mathcal{R}_{CF}\}$ curve with a single encoder and decoder using puncturing. We apply an identical random puncturing sequence $\underline{\pi}$ at nodes A and B to adjust the rate of the transmitted codewords $\underline{x}_A$ and $\underline{x}_B$. Let $p_{\pi}=\frac{|\pi|}{N}$ be the probability that a given $x_{A,B}[n]$ is punctured. The positions of the punctured bits are selected uniformly at random. The decoder treats the punctured bits as type $1$ messages. The effective channel can therefore be defined by
\begin{equation}
\underline{P}_{ch,\pi} = \left[\begin{array}{c}
p_1(\epsilon)(1-p_{\pi}) + p_{\pi} \\
p_2(\epsilon)(1-p_{\pi}) \\
p_3(\epsilon)(1-p_{\pi}) \\
p_4(\epsilon)(1-p_{\pi}) \\
p_5(\epsilon)(1-p_{\pi}) \\
\end{array}\right].
\end{equation}
The rate of a channel code is defined as the number of message bits $K$ divided by the codeword length $N$. Thus the rate of a code after puncturing is given by
\begin{equation}
\mathcal{R}_{\pi} = \frac{K}{N(1-p_{\pi})} = \mathcal{R}\frac{1}{1-p_{\pi}}.
\end{equation}
The puncture adjusted rate $\mathcal{R}_{\pi}$ is plotted as a function of $\epsilon_{thresh}$ in Fig. \ref{fig:dvdcLwBoth}. The punctured ensemble parameters are $(d_v,d_c,L,w)=(9,10,10000,100)$ which is a low rate ensemble. The type distribution evolution algorithm in \eqref{eq:TDE_dvdcLw} was allowed to run for $\ell_{max}=400,000$ iterations. Notice again that for each $\epsilon_{thresh}$, the achievable rate is closely upper bounded by $\max\{\mathcal{R}_{DF},\mathcal{R}_{CF}\}$. This means that the whole known equal rate region can be achieved using a single encoder and decoder at each transceiver.


\section{Conclusion}
We have numerically shown for some channel classes that JCF message passing decoding performs as well as DF and CF. We have derived a flexible class of channels for prototyping code designs. We have provided exact density evolution analysis for LDPC code ensembles for this class of TWEMACs. The universality of spatially coupled codes allows a single encoder and decoder to achieve any currently known equal exchange rate. This is an exciting problem which deserves further investigation and rigorous analysis.

\bibliographystyle{ieeetr}
\bibliography{lattice}

\end{document}